\documentclass[12pt,a4paper,epsf]{article}
\usepackage{graphics}
\usepackage{amssymb,amsmath}
\usepackage[dvips]{lscape,graphicx}
\usepackage{cite}
\usepackage{longtable}
\usepackage{hyperref}
\usepackage{graphics}
\usepackage{amssymb,amsmath}
\usepackage[dvips]{lscape,graphicx}
\newcommand{\ct}{\cite}
\newcommand{\lb}{\label}

\newcommand{\bc}{\begin{center}}
\newcommand{\ec}{\end{center}}
\newcommand{\bd}{\begin{displaymath}}
\newcommand{\ed}{\end{displaymath}}
\newcommand{\be}{\begin{equation}}
\newcommand{\ee}{\end{equation}}
\newcommand{\ba}{\begin{array}}
\newcommand{\ea}{\end{array}}
\newcommand{\bea}{\begin{eqnarray}}
\newcommand{\eea}{\end{eqnarray}}
\newcommand{\bt}{\begin{tabular}}
\newcommand{\et}{\end{tabular}}

\newcommand{\bp}{\begin{picture}}
\newcommand{\ep}{\end{picture}}
\newcommand{\bfi}{\begin{figure}}
\newcommand{\efi}{\end{figure}}

\def\fun#1#2{\lower3.6pt\vbox{\baselineskip0pt\lineskip.9pt
\ialign{$\mathsurround=0pt#1\hfil##\hfil$\crcr#2\crcr\sim\crcr}}}
\begin{document}
\vspace{1cm}
\title{ \bf{Standard Model and Graviweak Unification with (Super)Renormalizable Gravity.\\\vskip 0.3cm Part I:
Visible and Invisible Sectors of the Universe}}
\author{\large \bf L.V.~Laperashvili
${}^{1}$ \footnote{laper@itep.ru}\,\,,
H.B.~Nielsen${
}^{2}$ \footnote{hbech@nbi.dk}\,\, and A.~Tureanu ${}^{3}$
\footnote{anca.tureanu@helsinki.fi}\\[5mm]
\itshape{\large ${}^{1}$The Institute of Theoretical and
Experimental Physics,}\\
{\it National Research Center "Kurchatov Institute", }\\
{\it Bolshaya Cheremushkinskaya, 25, 117218 Moscow, Russia,}\\
\itshape{\large ${}^{2}$The Niels Bohr Institute, }\\
{\it Blegdamsvej 17-21, DK-2100 Copenhagen,
Denmark}\\
\itshape{\large ${}^{3}$Department of Physics, University of Helsinki,}\\{\it P.O. Box 64, FIN-00014 Helsinki, Finland}}
\date{}
\maketitle
\thispagestyle{empty}

\begin{abstract}
We develop a self-consistent $Spin(4,4)$-invariant model of the unification of gravity with weak $SU(2)$ gauge and Higgs
fields in the visible and invisible sectors of our Universe. We consider a general case of the graviweak unification, including the
higher-derivative super-renormalizable theory of gravity, which is a unitary, asymptotically-free and perturbatively consistent theory of the
quantum gravity.

\end{abstract}



\thispagestyle{empty}
\clearpage\newpage

\section{Introduction}

In the paper \ct{1}, using a general model \ct{2} of the
unification of gravity with $SU(N)$ or $SO(N)$ gauge and Higgs
fields,  we presented a model of unification of gravity with weak
$SU(2)$ gauge and Higgs fields, developing the ideas of
Ref.~\ct{3}. We have considered a $Spin(4,4)$-group of the
graviweak unification, which is spontaneously broken into the
$SL(2,C)^{(grav)}\times SU(2)^{(weak)}$.
In contrast to Ref.~\ct{1}, the main result of this investigation
(presented in Part I and Part II) is a case of the graviweak
unified model with 
super-renormalizable gravity. Such a model is constructed in 
agreement with experimental and astrophysical results.
In the present paper we present the embedding of the Standard
Model (SM) families into the group of TOE-unification existing at
the Planck era. We assume that at the early stage of the evolution
of the Universe the TOE-group (for example, $E_8$-group) is broken
down (say, in $\sim 10^{-43}$ sec after the Big Bang) to the
direct product of the gauge groups of internal symmetry and the
spacetime Lorentz group:
$$G_{TOE}\to G_{(Graviweak)}\times U(4)\to SL(2,C)^{(grav)}\times SU(2)^{(weak)}
\times U(4)$$ $$ \to SL(2,C)^{(grav)}\times SU(3)_{(color)}\times
SU(2)^{(weak)} \times U(1)_Y\times U(1)_{(B-L)} $$ $$\to
SL(2,C)^{(grav)}\times SU(3)_{(color)}\times SU(2)^{(weak)} \times
U(1)_Y. $$ Thus, below the sea-saw scale ($M_R\sim 10^{9}-
10^{14}$ GeV) we have the SM-group of symmetry:
$$ G_{SM} = SL(2,C)^{(grav)}\times SU(3)_{(color)}\times SU(2)^{(weak)} \times
U(1)_Y. $$ Spinors appear in multiplets of the gauge groups. Here it is necessary to emphasize that in our graviweak unified model there exists
a large breaking scale for  supersymmetry (more than $10^{18}$ GeV). Also we assume the existence in Nature of the invisible (mirror, or
hidden) world parallel to the visible Ordinary World (OW) \ct{2r,3r,4r,5r}. In our present paper this hidden sector of the Universe is a Mirror
World (MW) with broken Mirror Parity (MP) (see Refs.~\ct{7r,8r,9r,11r,11ar} and reviews \ct{12r,13r,16r}). The first part (Part I) of our
investigation "Standard Model and Graviweak Unification" is devoted to the main idea of Plebanski to describe gravity by connections and tetrads
as independent variables, to the problem of the hidden sector of our Universe and to the calculation of the action of the graviweak unification
with (super)renormalizable gravity. The second part - Part II - is devoted to  the main properties of the graviweak unified model with
(super)renormalizable gravity -- asymptotic freedom and the problem of unitarity, calculation of all coupling constants of theory and
consideration of their flow equations. Also the difference of coupling constants in the ordinary and hidden sectors is discussed. The paper is
organized as follows.  Section 2 is devoted to the visible and invisible sectors of our Universe. In Subsection 2.1
we consider the existence of the Mirror World (MW), which is a mirror copy of the Ordinary World (OW) and contains the same particles and types
of interactions as our visible world. In Subsection 2.2 we assume that our Universe has a mirror (hidden) world with a broken Mirror Parity
(MP): the Higgs VEVs of the visible and invisible worlds are not equal {\footnote {In this paper the superscript 'prime' denotes the M- (or
hidden H-) world.}, $\langle\phi\rangle=v, \quad \langle\phi'\rangle=v' \quad {\rm {and}}\quad v\neq v'$. The parameter characterizing the
violation of the MP is $\zeta = {v'}/{v} \gg 1$. We are using the estimate $\zeta\simeq 100$. In Section 3 we introduce the main ideas of the
Plebanski's theory of gravity. Section 4 is devoted to the problem of the existence of the mirror (hidden) world in Nature. In Sections 5 and 6
we constructed a model of the unification of super-renornalizable gravity with weak $SU(2)$ gauge and Higgs fields. This GWU model, including the
higher-derivative super-renormalizable theory of gravity is unitary, asymptotically-free and perturbatively consistent theory of the quantum
gravity. We considered the renormalization group flow of the higher derivative gravity in the 1-loop approximation, i.e. the RGE for the running
of the super-renormalizable gravitational coupling constants, predicted by our graviweak unification model. The theory is described by the overall
unification parameter $g_{uni}$. It was shown that this self-consistent GWU exists only at the high (Planck) scale. The graviweak unification
model in both, left-handed and right-handed (visible and invisible) sectors of the Universe, was considered in Subsection 5.4, where we
presented the actions for both sectors of the Universe, visible and mirror. Section 7 contains a summary and conclusions.

\section{Visible and invisible sectors of our Universe}

\subsection{Mirror World}

Refs.~\ct{2r,3r,4r,5r,7r,8r,9r,11r,11ar,12r,13r,16r} suggest the
hypothesis of the existence in Nature of the invisible mirror (or
hidden) world -- parallel to the visible (ordinary) one. The
Mirror World (MW) is a mirror copy of the Ordinary World (OW) and
contains the same particles and types of interactions as our
visible world. The observable elementary particles have 
left-handed (V-A) weak interactions, which violate P-parity. If a
hidden MW exists, then mirror particles participate in the
right-handed (V+A) weak interactions and have the opposite
chirality. Lee and Yang were the first \ct{2r} to suggest such a
duplication of the worlds, which restores the left-right symmetry
of the Nature. The term "Mirror Matter" was introduced by
Kobzarev, Okun and Pomeranchuk \ct{3r}. They first suggested the
MW as the hidden (invisible) sector of the Universe, which
interacts with the ordinary (visible) world only via gravity, or
another very weak interaction. They have investigated a variety of
phenomenological implications of such parallel worlds (see reviews
\ct{12r,13r}).
The SM group of symmetry $G_{SM}$ was enlarged to $G_{SM}\times
G'_{SM'}$, where $G_{SM}$ stands for the observable SM, while
$G'_{SM'}$ is its mirror gauge counterpart. These different worlds
are coupled only by gravity, or another very weak interaction
\ct{2r,3r,4r,5r}.
Superstring theory also predicts that there may exist in the
Universe another form of matter -- hidden (or 'shadow') matter,
which only interacts with ordinary matter via gravity or another
gravitational-strength interactions \ct{4r,5r}. According to the
superstring theory, these two worlds, ordinary and hidden, can be
sometimes (not always) very different -- at least unless we start
the two worlds up with the same quantum numbers. In general case,
we must distinguish initial conditions and the equations of
motion. These two worlds can be viewed as parallel branes in a
higher dimensional space, where visible particles are localized on
the one brane and hidden particles -- on the another brane, and
gravity propagates in the bulk.
In Refs.~\ct{41r,43r,46r} we considered the theory of the
superstring-inspired $E_6$ unification with different types of the
breaking of the $E_6$ symmetry in the visible and hidden worlds.

\subsection{Mirror world with broken mirror parity}

If the ordinary and mirror worlds are identical, then O- and
M-particles should have the same cosmological densities. But this
is immediately in conflict with recent astrophysical measurements
if the MW-density should be identified with Dark Matter.
Astrophysical and cosmological observations (see for example
\ct{47r,48r}) have revealed the existence of the Dark Matter (DM)
which constitutes about 25\% of the total energy density of the
Universe. This is five times larger than all the visible matter,
$\Omega_{DM}: \Omega_{M} \simeq 5 : 1$.  Mirror particles have
been suggested as candidates for the inferred dark matter in the
Universe (see Refs.~\ct{7r,8r,9r,11r,11ar,12r,13r,16r}).
Therefore, the mirror parity (MP) is not conserved, and the OW and
MW are not identical. In Refs.~\ct{7r,8r,9r,11r} it was suggested
that the VEVs of the Higgs doublets $\phi(=H)$ and $\phi'(=H')$
are not equal:
\be \langle\phi\rangle=v,\quad \langle\phi'\rangle=v' \quad {\rm
{and}}\quad v\neq v'.
                          \lb{39M} \ee
The parameter characterizing the violation of MP:
\be  \zeta = \frac {v'}{v} \gg 1
                          \lb{40M} \ee
was introduced and estimated in Refs.~\ct{7r,8r,9r,11r} and
\ct{5y,2y,3y,4y}:
\be  \zeta > 30,\quad \zeta \sim 100.
                          \lb{41M} \ee
Then the masses of mirror fermions and massive bosons are scaled
up by the factor $\zeta$ with respect to the masses of their
OW-counterparts:
\be m'_{q',l'} = \zeta m_{q,l},
                          \lb{42M} \ee
and \be
   M'_{W',Z',\phi'} = \zeta  M_{W,Z,\phi}, \lb{43M} \ee
while photons and gluons remain massless in both worlds.
In the language of neutrino physics, the O-neutrinos
$\nu_e,\,\,\nu_{\mu},\,\,\nu_{\tau}$ are active neutrinos, and the
M-neutrinos $\nu'_e,\,\,\nu'_{\mu},\,\,\nu'_{\tau}$ are sterile
neutrinos \ct{5y}. If MP is conserved ($\zeta = 1$), then the
neutrinos of the two sectors are strongly mixed (see Refs.~
\ct{7r,8r,9r,11r}). However, the present experimental and
cosmological limits on the active-sterile neutrino mixing do not
confirm this result. The 'neutrino-mirror neutrino' oscillations
were investigated in Refs.~\ct{6yy,1y,5y,5yy}.  In
Refs.~\ct{6y,7y} the exact parity symmetry explains the solar
neutrino deficit, the atmospheric neutrino anomaly and the LSND
experiment.
In the context of the SM, in addition to the fermions with
non-zero gauge charges, one introduces also the gauge singlets,
the so-called right-handed neutrinos $N_a$ with large Majorana
mass terms. They have equal masses in the OW  and MW \ct{7r,8r}:
\be M'_{\nu,a} = M_{\nu,a}. \lb{44M} \ee
According to the usual seesaw mechanism \ct{8y}, heavy
right-handed neutrinos are created in the OW at the seesaw scale
$M_R\sim 10^9-10^{14}$ GeV.

\section{Gravity in the Plebanski's formulation
of General Relativity}

Originally General Relativity (GR) was formulated by Einstein as
the dynamics of a metric, $g_{\mu\nu}$. Later Plebanski \ct{17r},
Ashtekar \ct{18r,19r} and other authors \ct{20r,21r} presented GR
in a self-dual approach, in which the true configuration variable
is a self-dual connection corresponding to the gauging of the
local Lorentz group, $SO(1,3)$, or the spin group, $Spin(1,3)$. In
the unification models \ct{1,2}, the fundamental variable is a
connection, $A$, valued in a Lie algebra, $\mathfrak g$, that
includes a subalgebra $\tilde {\mathfrak  g}$:
\be \tilde {\mathfrak  g} = {\mathfrak  g}^{(spacetime)} \oplus
{\mathfrak g}_{YM}, \lb{1} \ee
which is the direct sum of the Lorentz algebra and a Yang--Mills gauge algebra. Previously  graviweak and gravi-electro-weak unified models were
suggested in Ref.~\ct{23r,24r,40r}. The gravi-GUT unification was developed in \ct{25r,27r,28r,29r}. In the Plebanski's formulation of the
4-dimensional theory of gravity \ct{17r}, the gravitational action is the product of two 2-forms (see \ct{17r,18r,19r,20r,21r} and
\ct{30r,31r,34r,37r,39r}), which are constructed from the connections $A^{IJ}$ and tetrads (or frames) $e^I$ considered as independent dynamical
variables. Both $A^{IJ}$ and $e^I$ are 1-forms:
\be   A^{IJ} =  A_{\mu}^{IJ}dx^{\mu} \quad {\mbox{and}}\quad
       e^I = e_{\mu}^Idx^{\mu}.
                          \lb{2} \ee
The indices $I,J = 0,1,2,3$ refer to the spacetime with Minkowski
metric $\eta_{IJ}$: $\eta^{IJ} = {\rm diag}(1,-1,-1,-1)$. This is
a flat space which is tangential to the curved space with the
metric $g_{\mu\nu}$. The world interval is represented as $ds^2 =
\eta_{IJ}e^I \otimes e^J$, i.e.
\be g_{\mu\nu} = \eta_{IJ} e^I_{\mu}\otimes e^J_{\nu}.
 \lb{4} \ee
Considering the case of the Minkowski flat spacetime with the
group of symmetry $SO(1,3)$, we have the capital latin indices
$I,J,...=0,1,2,3$, which are vector indices under the rotation
group $SO(1,3)$.
The 2-forms $B^{IJ}$ and $F^{IJ}$ are defined as:
\be  B^{IJ} = e^I\wedge e^J = \frac 12
e_{\mu}^Ie_{\nu}^Jdx^{\mu}\wedge dx^{\nu},\qquad F^{IJ} = \frac 12
F_{\mu\nu}^{IJ}dx^{\mu}\wedge dx^{\nu}. \lb{5} \ee
Here the tensor $F_{\mu\nu}^{IJ}$ is the field strength of the
spin connection $A_{\mu}^{IJ}$:
\be
     F_{\mu\nu}^{IJ} = \partial_{\mu}A_{\nu}^{IJ} -
         \partial_{\nu}A_{\mu}^{IJ} + [A_{\mu}, A_{\nu}]^{IJ},
                               \lb{7} \ee
which determines the Riemann--Cartan curvature:
\be
       R_{\kappa \lambda \mu \nu} = e_{\kappa}^I e_{\lambda}^JF_{\mu\nu}^{IJ}.
                  \lb{8} \ee
In the Plebanski BF-theory, the gravitational action with nonzero
cosmological constant $\Lambda$ is given by the integral:
\be  I_{GR} = \frac{1}{\kappa^2}\int
\epsilon^{IJKL}\left(B^{IJ}\wedge
    F^{KL} + \frac{\Lambda}{4}B^{IJ}\wedge B^{KL}\right),
                                  \lb{11} \ee
where $\kappa^2=8\pi G_N$, $G_N$ is the gravitational constant, $
M_{Pl}^{red.} = 1/{\sqrt{8\pi G_N}}$.
For any antisymmetric tensors $F_{\mu\nu}$ and $A^{IJ}$ there
exist dual tensors given by the Hodge star dual operation:
\be
 F^*_{\mu\nu}\equiv \frac {1}{2\sqrt{-g}}{\epsilon
 _{\mu\nu}} ^{\rho\sigma}F_{\rho\sigma},\qquad A^{\star IJ} = \frac 12
{\epsilon^{IJ}}_{KL}A^{KL}.
                                 \lb{12} \ee
Here $\epsilon$ is the completely antisymmetric tensor with
$\epsilon^{0123} = 1$.
We can define the algebraic self-dual (+) and anti-self-dual (-)
components of $A^{IJ}$:
\be A^{(\pm)\,IJ}=({\cal P}^{\pm}A)^{IJ} = \frac 12 (A^{IJ} \pm
iA^{\star\,IJ}),
                                 \lb{14} \ee
and as a result we have:
\be  A^{(+)} = -iA^{\star(-)},\qquad A^{(-)} = iA^{\star(+)},
                                  \lb{14a} \ee
or
\be   A^{\star(+)} = -iA^{(-)},\qquad A^{\star(-)} = iA^{(+)}.
                                  \lb{14b} \ee
The two projectors ${\cal P}^{\pm}= \frac 12(\delta^{IJ}_{KL}
\pm\frac i2 \epsilon^{IJ}_{KL})$ realize explicitly the familiar
homomorphism: \\
\be
   SO(1,3)_C = SL(2,C)_L \otimes SL(2,C)_R,  \lb{15a} \ee
or
\be
   \mathfrak{so}(1,3)_C = \mathfrak{sl}(2,C)_L \oplus
   \mathfrak{sl}(2,C)_R,
                                 \lb{15b} \ee
which rather than self-dual (+) and anti-self-dual (-) are more
commonly dubbed right-handed (R) and left-handed (L).
If we consider a Wick rotation in the gravitational theory,
replacing the time $t$ by imaginary time $t'=it$, then we obtain
the gravity in the Euclidean spacetime with $SO(4)$-group of
symmetry (see for example \ct{34r}), and instead of
Eqs.~(\ref{15a}) and (\ref{15b}), we have:
\be
   SO(4)= SU(2)_L \otimes SU(2)_R,  \lb{15c} \ee
or
\be
   \mathfrak{so}(4) = \mathfrak{su}(2)_L \oplus
   \mathfrak{su}(2)_R.
                                 \lb{15d} \ee
The self-dual and anti-self-dual tensors $A^{(\pm)IJ}$ have only
three independent components given by $IJ = 0i$, where $i = 1, 2,
3$. To make the mapping more explicit, it is convenient to pick
out the time direction equal to zero, and define:
\be
                 A^{(\pm)i} = \pm 2A^{(\pm) 0i}
                     \lb{16} \ee
with $i = 1,2,3$.
The correct gauge was provided by Plebanski, when he introduced in
the gravitational action the Lagrange multipliers $\psi_{ij}$ --
an auxiliary fields, symmetric and traceless. These auxiliary
fields provide a correct number of constraints, and we obtain the
following gravitational action:
\be I_{gravity}(\Sigma,A,\psi) = \frac{1}{\kappa^2} \int
[\Sigma^i\wedge F^i +
 (\Psi^{-1})_{ij}\Sigma^i\wedge \Sigma^j].
                      \lb{17} \ee
The usual notations:
\be \Sigma^i=2B^{0i},   \lb{18} \ee
and
\be (\Psi^{-1})_{ij} = \psi_{ij} - \frac{\Lambda}{6}\delta_{ij}
\lb{19} \ee
are presented in action given by Eq.~(\ref{17}).
Following the ideas of Ref.~\ct{3}, we distinguish the two worlds
of the Universe, visible and invisible, and consider the two
sectors of gravity: left-handed gravity and right-handed gravity.
The self-dual left-handed gravity is described by the following
action:
\be I^{(grav)}_{(OW)}(\Sigma^{(+)},A^{(+)},\psi) =
\frac{1}{\kappa^2} \int [\Sigma^{(+)i}\wedge F^{(+)i} +
 (\Psi^{-1})_{ij}\Sigma^{(+)i}\wedge \Sigma^{(+)j}].
                      \lb{20} \ee
Using the simpler self-dual variables instead of the full Lorentz
group, Plebanski \ct{17r} and the authors of
Refs.~\ct{18r,19r,20r,21r} suggested to consider in the visible
sector of our Universe the left-handed
$\mathfrak{sl}(2,C)^{(grav)}_L$-invariant gravitational action
(\ref{20}) with self-dual fields $F^{(+)i}$ and $\Sigma^{(+)i}$.
If there exists in Nature a duplication of worlds with opposite
chiralities �- Ordinary and Mirror -- we can consider the
left-handed gravity in the Ordinary world and the right-handed
gravity in the Mirror world.
The anti-self-dual right-handed gravitational action of the mirror
world MW is given by the following integral:
\be I^{(grav)}_{(MW)}(\Sigma^{(-)},A^{(-)},\psi') =
\frac{1}{{\kappa'}^2} \int [\Sigma^{(-)i}\wedge F^{(-)i} +
 ({\Psi'}^{-1})_{ij}\Sigma^{(-)i}\wedge \Sigma^{(-)j}].
                      \lb{21} \ee
In Eqs.~(\ref{20}) and (\ref{21}) we have:
\be  \Sigma^{(\pm) i} = e^0\wedge e^i \pm i\frac 12
     \epsilon^i_{jk}e^j\wedge e^k.
                          \lb{22} \ee
The self-dual action (\ref{20}) is equivalent to the
Einstein-Hilbert action for general relativity with the Einstein's
cosmological constant $\Lambda$ \ct{17r}:
\be I_{EH} = - \frac{1}{{\kappa}^2} \int d^4x\sqrt{-g}(\frac 12 R
- \Lambda),   \lb{23} \ee
where $R$ is a scalar curvature.
A problem of constraints in the Plebanki's theory of GR was in
detail studied in Refs.~\ct{21r,22r} (see also \ct{30r,31r,34r}).
Plebanski considered the action (\ref{20}) with the following
constraints:
\be  \Sigma^{(+)i}\wedge \Sigma^{(+)j} - \frac 13
\delta^{ij}\Sigma^{(+)k}\wedge \Sigma^{(+)}_k = 0,   \lb{24} \ee
and
\be  \Sigma^{(+)i}\wedge \Sigma^{(-)j} = 0.  \lb{25} \ee
The variables $\Sigma_{\mu\nu}^i$ have 18 degrees of freedom. The
five conditions (\ref{24}) leave 13 degrees of freedom, and the
condition (\ref{25}) leaves 10 degrees of freedom, which coincides
with a number of degrees of freedom given by the metric tensor
$g_{\mu\nu}$. This circumstance confirms the equivalence of the
actions (\ref{20}) and (\ref{23}). And now we have the following
groups describing GR:
\be
   \mathfrak{so}(1,3)_C = \mathfrak{sl}(2,C)_L \oplus
   \mathfrak{sl}(2,C)_R.
                                 \lb{26} \ee
If the anti-self-dual right-handed gravitational world is absent
in Nature ($\Sigma^{(-)} = 0$ and $F^{(-)} = 0$ ), then the
gravity of our world, in which we live, is presented only by the
self-dual left-handed Plebanski action (\ref{20}) equivalent to
the Einstein-Hilbert's gravity (\ref{23}). This is a main
assumption of Plebanski. The same self-dual formulation of General
Relativity was developed later by Ashtekar \ct{18r,19r}.
In Section 4 and below we wish to use the following notations for
$X=A,B,F,\Sigma$:
$$ X = X^{(+)}, \qquad X' = X^{(-)}.$$

\section{Does the Mirror or Hidden world of the Universe really
exist?}

Here we can consider four possibilities.

I) The mirror right-handed world is absent in the Universe: the
mirror particles and the mirror gravity do not exist in Nature,
what means that $\Sigma' = \Sigma^{(-)} = 0$ and the right-handed
connection $A'= A_R = A^{(-)} = 0$.

II) The mirror (or hidden) right-handed world is separated from
the visible (ordinary) left-handed one,  $\Sigma = \Sigma^{(+)}
\neq 0$, $\Sigma' = \Sigma^{(-)} \neq 0$, and the connections $A =
A_L = A^{(+)}$, $A'= A_R =A^{(-)}$ are not zero, but the ordinary
and mirror gravitational fields do not interact directly.
There are several fundamental ways by which the mirror (hidden)
world can communicate with our visible world. Two worlds can
interact via a very weak interaction of the singlet scalar fields.
It was shown in Ref.~\ct{33x} that an ultralight scalar field with
mass around $10^{-22}- 10^{-23}$ eV is a viable dark mater
candidate, and this field can be detected by the planned SKA
pulsar timing array experiments (see \ct{34x}), which considered
the gravitational field of the galactic halo composed  of such
dark matter.
The authors of Ref.~\ct{10y} suggested that there exists an
interaction between the ordinary and mirror Higgs doublets, $\phi$
and $\phi'$, respectively:
\be L_H = \lambda_1(\phi^{\dagger} \phi ) ({\phi'}^{\dagger}
\phi'). \lb{45M} \ee
In Ref.~\ct{11y} the existence of the additional interaction:
\be L_Y = \epsilon_Y F_{Y,\mu\nu}{F'_{Y}}^{\mu\nu} \lb{46M} \ee
was assumed, where $F_{Y,\mu\nu}$ and  $F'_{Y,\mu\nu}$ are the
$U(1)_Y$ and $U(1)'_Y$ field strength tensors, respectively.

Also the right-handed  Majorana neutrinos $N_a$ can communicate
between visible and hidden worlds (see for example \ct{12y,13y}).

III) (a) The third possibility concerns the theory of the strong
mixing of $g_{\mu\nu}^{L,R}$ -- the so called "bigravity theory"\,
(see for example \ct{14y}). It was assumed in \ct{14y} that
ordinary matter and mirror matter interact with two separate
metric tensors $g^L_{\mu\nu}$ and $g^R_{\mu\nu}$, i.e. each sector
has its own GR-like gravity. The effective action of this model
contains Einstein-Hilbert terms in each sector and a mixing term
between the two sectors:
\be I = \int d^4x[\sqrt{g^L}(\frac{M_{Pl}}{2}R^{(L)} + L_1) +
\sqrt{g^R}(\frac{M_{Pl}}{2}R^{(R)} + L_2) +
{(g_Lg_R)}^{1/4}L_{mix}], \lb{5z} \ee
where $L_1$ and $L_2$ are the Lagrangians respectively for the
ordinary and mirror particles/fields, and $L_{mix}$ again
describes possible interaction terms between the ordinary and
mirror worlds.

(b) The mixing of $g_{\mu\nu}^{L,R}$ can be so strong that the
left-handed gravity coincides with the right-handed gravity: the
left-handed and right-handed connections are equal, $A=A'$, i.e.
$A_L=A_R$, what means that a dual part $A^\star$ of the connection
is zero. In this case $g_{\mu\nu}^L = g_{\mu\nu}^R$, and the
left-handed and right-handed gravity equally interact with visible
and mirror matters.
To describe the real world we have to restrict the solutions of
theory to those in which the metric is real. In spite of the fact
that the metric is not a fundamental field in the Plebanski
action, we can specify the modified reality conditions. For this
purpose, it is convenient to consider the two component spinor
indices (see for example \ct{37r}): $a,b = 0,1$ are left handed
spinor indices, while $a',b' = 0',1'$ are right handed spinor
indices. This allows us to easily distinguish the left and right
handed fields. The connection decomposes into:
\be A^{IJ} = A^{aa'bb'} = \epsilon^{ab}A^{a'b'} + A^{ab}
\epsilon^{a'b'}, \lb{47M} \ee
and the two forms $B^{IJ}$ similarly decompose. The remarkable
fact is that the constructed metric is cubic in $B$-fields
\ct{21r,40r}. In fact, two metrics can be built, out of the left
and right parts of $B$ (or $\Sigma$), which are called the left
and right Urbantke metrics \ct{21r,40r}:
\be  g_{\mu\nu}^L =
\epsilon^{\alpha\beta\gamma\delta}B^a_{\mu\alpha
\,b}B^b_{\nu\beta\,c}B^c_{\gamma\delta\,a},     \lb{1z} \ee
\be  g_{\mu\nu}^R =
\epsilon^{\alpha\beta\gamma\delta}{B'}^{a'}_{\mu\alpha
\,b'}{B'}^{b'}_{\nu\beta\,c'}{B'}^{c'}_{\gamma\delta\,a'}, \lb{2z}
\ee
in which $\epsilon^{\alpha\beta\gamma\delta}$ is the Levi-Civita
symbol.
In the Minkowski spacetime background, in low energy limit, the
metric can be expanded by the Feynman expansion:
\be  g_{\mu\nu}^{L,R} = \eta_{\mu\nu} + \kappa_{L,R}
h_{\mu\nu}^{L,R}, \lb{3z} \ee
where $h^{L,R}_{\mu\nu}$ are left- and right-handed gravitons,
respectively.
Considering the interaction between ordinary and mirror worlds, we
can discuss a phenomenology of the two gravitons
$h^{L,R}_{\mu\nu}$. Then a parity even combination:
\be     h_{\mu\nu} = h^L_{\mu\nu} + h^R_{\mu\nu}     \lb{4z} \ee
will be massless due to the diffeomorphism invariance, and will
correspond to the standard graviton. This graviton couples
universally to L- and R- matters.
On the other hand, the parity odd combination:
\be   \tilde h_{\mu\nu} = h^L_{\mu\nu} - h^R_{\mu\nu}     \lb{5z}
\ee
is not protected by diffeomorphisms and may be massive. If the parity odd graviton $\tilde h$ has the Planck-scale mass, then it would be
unobservable at low energy. In this case, there exists the overall gravity in the Universe, and the ordinary (visible) world interacts with the
mirror (hidden) world via this gravity. Theory III) depends on details of the full theory at energy beyond the Planck scale. If the parity odd
graviton $\tilde h_{\mu\nu}$ is sufficiently light, then it would give rise to the polarization-dependent gravitational effects during the
detection of the gravitational waves in the CMB or pulsar timing programs. Then the parity violation is applied not only to the weak
interaction, but also to the gravitational sector. The parity violation of gravity was considered in Ref.~\ct{15y}, which proposed a test of
this effect through coincident observations of gravitational waves and short gamma-ray bursts from binary mergers involving neutron stars. Such
gravitational waves are highly left or right circularly-polarized due to the geometry of the merger. Using localization information from the
gamma-ray burst, ground-based gravitational wave detectors can measure the distance to the source with reasonable accuracy. Gravitational parity
violation would manifest itself as a discrepancy between the distance measurements. The effective theory, leading to such gravitational
parity-violation, is the Chern-Simons theory of gravity \ct{15yy}. The future experiments detecting the parity non-conservation in gravity are
planning in the framework of the development of the future investigations of CMB \ct{34sr,35sr,36sr,39sr}. If our Universe has chosen the
picture III, then the dynamics of the total Universe, visible and invisible, is governed by the following action:
\be   I = \int d^4x( L_{(grav)} + L_{SM} + L'_{SM'} + L_{(mix)}),
\lb{48M} \ee
where $L_{(grav)}$ is the gravitational (Einstein-Hilbert) low-energy Lagrangian, describing gravity in both (O- and M-) worlds, $L_{SM}$ and
$L'_{SM'}$ are the Standard Model Lagrangians in the O- and M-worlds, respectively, $L_{(mix)}$ is the Lagrangian describing all mixing terms,
which give very small contributions to physical processes: mirror particles have not been seen so far, and the communication between visible and
hidden worlds is hard. Searching for mirror particles at the LHC was discussed in Ref.~\ct{16y}. We can imagine that a fraction of the mirror
matter exists in the form of mirror galaxies, mirror stars, mirror planets, etc. (see for example Refs.~\ct{17y,18y,19y}). These objects can be
detected by the gravitational microlensing methods \ct{18y}. Such researches show fascinating results (see for example \ct{13r} and \ct{16r}). A
Nature of our Universe can be understood by future experiments with CMB, similar to WMAP, "Hubble", "Planck", "BICEP2" \ct{34sr,35sr,36sr,39sr},
and also depends on the future LHC results.

\section{Graviweak unified model with renormalizable gravity}

Developing the graviweak unification model in the visible sector
of the Universe, we started in Ref.~\ct{1} with a $\mathfrak g =
\mathfrak {spin}(4,4)$-invariant extended Plebanski's action:
\be I(A, B, \Phi) = \frac{1}{g_{uni}} \int_{\mathfrak M}\langle BF
+ B\Phi B + \frac 13 B\Phi \Phi \Phi B \rangle. \lb{27} \ee
The wedge product $\langle...\rangle$ is assumed between the
forms. The action (\ref{27}) contains a parameter of the
unification $g_{uni}$. The connection, $A =\frac 12
A^{IJ}\gamma_{IJ}$, is an independent physical variable describing
the geometry of the spacetime, while $B$ and $\Phi$ are considered
as auxiliary fields \ct{2}. All 2-forms, $F=\frac 12
F^{IJ}\gamma_{IJ}$ and $B=\frac 12 B^{IJ}\gamma_{IJ}$, are
$\mathfrak{spin}(4,4)$-valued fields. Here $F = dA + \frac 12 [A,
A]$. The fields $F^{IJ}$ and $B^{IJ}$ again are given by
Eqs.~(\ref{5}) and (\ref{7}), but now the indices $I,J$ run over
all $8\times 8$ values: $I,J = 1,2,...,7,8$ ($I,J=1,5,6,7$ -
timelike components, and $I,J=2,3,4,8$ - spatial ones). The
auxiliary field $\Phi$ is a symmetric linear operator, which
transforms bivectors to bivectors and 2-forms to 2-forms, it has
the indices ${{\Phi_{\mu\nu}}^{\rho\sigma IJ}}_{KL}$. As an
example, the second term of the action (\ref{27}) is
\be  \langle B\Phi B \rangle = \frac
1{32}\epsilon^{\mu\nu\rho\sigma}B_{\mu\nu
IJ}{{\Phi_{\rho\sigma}}^{\varphi\chi IJ}}_{KL}
{B_{\varphi\chi}}^{KL}d^4x. \lb{28} \ee
The field equations obtained by varying the fields $A,B$ and
$\Phi$  are:
\be  {\cal D}B = dB + [A,B] = 0,  \lb{29} \ee
where ${\cal D}$ is the covariant derivative, ${\cal D}_{\mu}^{IJ}
= \delta^{IJ}\partial_{\mu} - A_{\mu}^{IJ}$,
\be   F = -2\left(\Phi +\frac 13\Phi \Phi \Phi\right)B,  \lb{30}
\ee
and
\be   \frac 1{32}\epsilon^{\mu\nu\rho\sigma} B_{\mu\nu IJ}
{B_{\varphi\chi}}^{KL} = - \frac 1{512}\epsilon^{\mu\nu\rho\sigma}
B_{\mu\nu IJ} {{\Phi_{\varphi\chi}}^{\psi\omega
KL}}_{MN}{{\Phi_{\psi\omega}}^{\xi\zeta
MN}}_{PQ}{B_{\xi\zeta}}^{PQ}. \lb{31} \ee
The first equation (\ref{29}) describes the dynamics, while
Eqs.~(\ref{30}) and (\ref{31}) determine the auxiliary fields $B$
and $\Phi$.
The specific action (\ref{27}) with the "Mexican hat" potential
for $\Phi$ has been chosen here in accordance with Ref.~\ct{2},
because such a choice allows symmetry breaking to a non-trivial
vacuum expectation value (VEV). Of course, more general actions
with arbitrary potential, $U(\Phi)$, can also be chosen for
another type of unification. But here we consider the action
(\ref{27}), which leads to a simple analysis.

\subsection{Symmetry breaking}

According to Refs. \ct{1} and \ct{2}, we can present the following
spontaneous symmetry breaking of the $\mathfrak g$-invariant
action (\ref{27}):
\be   \tilde {\mathfrak g} = {{\mathfrak sl}(2,C)}^{(grav)}_L
\oplus {\mathfrak su}(2)_L. \lb{32} \ee
Below the indices $\large a, b \in \{0,1,2,3\}$ are used to sum
over a subset of $I, J \in {1,2, ...,7,8}$ for $I,J=1,2,3,4$, and
thereby select a $\mathfrak {spin}(1,3)$ subalgebra of $\mathfrak
{spin}(4,4)$. The indices $m,n \in \{5,6,7,8\}$ sum over the rest.
We also consider $\large i, j \in \{1,2,3\}$, thus selecting
$\mathfrak {sl}(2,C)_{L,R}^{grav}$ subalgebras of $\mathfrak
{spin}(4,4)$.
The equations of motion (\ref{29})-(\ref{31}) help us to obtain a
symmetry breaking ansatz for $\Phi$ (of course, phenomelogically
assuming no breaking of Lorentz invariance) \ct{2}:
\be {{\Phi_{\mu\nu}}^{\rho\sigma ab}}_{cd} =
a_1\delta_{\mu\nu}^{\rho\sigma}\delta^{ab}_{cd } + b_1(e_{\mu}^f)
(e_{\nu}^g){\epsilon_{fg}}^{kl}(e^{\rho}_k)
(e^{\sigma}_l)\delta_{cd}^{ab} +
c_1\delta_{\mu\nu}^{\rho\sigma}{\epsilon^{ab}}_{cd} +
d_1{\epsilon_{\mu\nu}}^{\rho\sigma}{\epsilon^{ab}}_{cd}, \lb{33}
\ee
where $a_1,\, b_1,\, c_1,\, d_1$ are parameters determined by the
equations of motion.
Using the notations of Ref.~\ct{2} for Eq.~(\ref{33}), we have:
\be  \Phi = a_1 + b_1 \ast + c_1 \star + d_1\ast \star. \lb{34}
\ee
In the present paper, in contrast to Refs.~\ct{1} and \ct{2}, we
do not consider the first class of solutions for $\Phi$, when we
have $a_1=c_1=d_1=0$ and $b_1=1$. We investigate the second class
of solutions (see \ct{2}) for ansatz (\ref{33}) with any
non-trivial values of parameters. In this case, according to
Eq.~(\ref{30}), we have:
\be B = \Omega F, \lb{35} \ee
where
\be \Omega = - \frac 12\left(\Phi +\frac 13\Phi \Phi
\Phi\right)^{-1}. \lb{36} \ee
Using the ansatz (\ref{33}) (or (\ref{34})) for $\Phi$, it is easy
to obtain that $\Omega $ has the structure analogous to
(\ref{33}), (\ref{34}):
\be  \Omega = a_2 + b_2 \ast + c_2 \star + d_2\ast \star, \lb{37}
\ee
with new parameters $a_2,\,b_2,\, c_2,\,d_2$ determined by the
equations of motion.
Taking into account the equations of motion and the result
(\ref{35}), we can present the following action for the Graviweak
unification:
\be   I(A, \Omega) = \frac{1}{2g_{uni}} \int_{\mathfrak M} \langle
\Omega FF \rangle, \lb{38} \ee
where
\be  \langle \Omega F F \rangle =
\frac{d^4x}{32}\epsilon^{\mu\nu\rho\sigma}{{\Omega_{\mu\nu}}^{\varphi\chi
IJ}}_{KL}F_{\varphi\chi IJ} {F_{\rho\sigma}}^{KL}. \lb{38a} \ee
\subsection{Symmetry breaking to gravity}

The ansatz (\ref{35})-(\ref{38a}) for $\Omega$ distinguishes a
subalgebra $\tilde {\mathfrak  g}$ of the algebra ${\mathfrak  g}
= \mathfrak{spin}(4,4)$, and we can consider separate parts of our
connection $A$:
\be  A = \frac 12 A^{IJ}\gamma_{IJ} = \frac 12
(A^{ab}\gamma_{a}\gamma_{b} + A^{am}\gamma_{a}\gamma_{m} +
A^{ma}\gamma_{m}\gamma_{a} + A^{mn}\gamma_{m}\gamma_{n}),
                               \lb{39} \ee
or
\be A = \frac 12 \omega + \frac 14 E + A_W, \lb{40} \ee
where the gravitational spin connection is:
\be   \omega = \frac 12 \omega^{ab}\gamma_a\gamma_b, \lb{41} \ee
the frame-Higgs connection
\be   E = E^{am}\gamma_a\gamma_m   \lb{42} \ee
is valued in the off-diagonal complement of $\mathfrak
{spin}(4,4)$, and the weak gauge field is:
\be  A_W = \frac 12 A^{mn}\gamma_m\gamma_n . \lb{43} \ee
In Eqs.~(\ref{39})-(\ref{43}) the indices $\large a, b \in
\{0,1,2,3\}$ are used to sum over a subset of $I, J \in {1,2,
...,7,8}$ for $I,J=1,2,3,4$, and thereby select a $\mathfrak
{spin}(1,3)$ subalgebra of $\mathfrak {spin}(4,4)$. The indices
$m, n \in \{5,6,7,8\}$ sum over the rest, and further we use
$\tilde m, \tilde n = m-5,n-5$, i.e. $\tilde m, \tilde n \in
\{0,1,2,3\}$. Here $\gamma_a = \{\gamma_0, \vec {\gamma} \}$ and
$\gamma_{\tilde m} = \{{\bf 1}, \gamma_0 \vec {\gamma}\}$. We have
chosen
$\gamma_I=1,...,8=\{1_4,\gamma_0,\gamma_i,\gamma_0\gamma_i\}$, and
$\tilde m = 0,1,2,3$ correspond to $m=I=5,6,7,8$.

Since matrix $A^{IJ}$ is antisymmetric, we have antisymmetric
matrices $\omega^{ab}$ and $A_W^{\tilde m\tilde n}$, but (see
\ct{2}):
\be   E^{a\tilde m} = e^a_{\mu}\phi^{\tilde m} dx^{\mu},
                                                       \lb{44}                                                         \ee
where the fields $\phi^{\tilde m}$ are four components of the
complex scalar Higgs field.

\subsection{The Left and Right worlds of the Universe}

Developing the ideas of Refs.~\ct{1,3}, we distinguish the Left
and Right worlds of the Universe, considering the graviweak
unified model in both sectors of the Universe, visible and
invisible, which are described respectively by the following
connections:
\be A^{IJ} = A^{IJ}_L = A^{(+)IJ},  \lb{46} \ee
\be {A'}^{IJ} = A^{IJ}_R = A^{(-)IJ}.  \lb{47} \ee
In Ref.~\ct{3} we suggested to describe the gravity in the visible
Universe by the self-dual left-handed Plebanski's gravitational
action, while the gravity in the invisible Universe -- by the
anti-self-dual right-handed gravitational action.
As it was shown in Section 3, the best way to study many aspects
of the Lorentz group is via its Lie algebra. Since the Lorentz
group is SO(1,3), its Lie algebra is reducible and can be
decomposed into two copies of the Lie algebra:
\be {\mathfrak  so(1,3)_C} = {\mathfrak sl}(2,C)_L \oplus
{\mathfrak sl}(2,C)_R. \lb{48} \ee
In particle physics, a state that is invariant under one of these
copies of $SO(1,3)_C$ is said to have chirality, either
left-handed or right-handed, according to which copy of
$SO(1,3)_C$ it is invariant under.
Self-dual tensors transform non-trivially only under $SL(2,C)_L$
and are invariant under $SL(2,C)_R$. By this reason, they are
called "left-handed" tensors. Similarly, anti-self-dual tensors,
non-trivially transforming only under $SL(2,C)_R$, are called
�right-handed� tensors. These self-dual and anti-self-dual tensors
$\omega^{(\pm)ab}$ have only three independent components given by
$ab = 0i\,\, (i = 1,2,3)$:
\be \omega^{(\pm)i} = \pm 2\omega^{(\pm)0i}.  \lb{50} \ee
As a result, we have the Left world of the Universe, described by
the left-handed self-dual 1-form connections:
\be A = A_L = A^{(+)} = \frac 12 \omega + \frac 14 E + A_W,
\lb{51} \ee
and  the Right world of the Universe, described by the
right-handed anti-self-dual 1-form connections:
\be  A' = A_R = A^{(-)} = \frac 12 \omega' + \frac 14 E' +
A'_{W'}, \lb{52} \ee
where gauge fields $A^{(')}$ are $A$ in L-world and $A'$ in
R-world.
For the weak gauge sector we have:
\be  A^{(')}_W = \frac 12 A_W^{(')\tilde m \tilde n}\gamma_{\tilde
m}\gamma_{\tilde n}, \lb{53} \ee
which are valued in $\mathfrak su(2)_{L,R}$-algebra, respectively.
Then gauge fields $A_W^{(')\tilde m\tilde n}$ have only three
non-zero components:
\be A_W^{(')i} = \pm 2A_W^{(')0i}, \lb{54} \ee
and the vector fields $A_W^{(')i}$ ($i=1,2,3$) transform as
adjoint vectors under the corresponding weak $SU(2)_{L,R}$ gauge
group, respectively.

Taking into account the self- (or anti-self) duality, we see in the
ansatz (\ref{37}) the following equivalence: $$a_2 = c_2\star\quad
{\rm{and}}\quad b_2\ast = d_2\ast \star.$$ Then instead of
(\ref{37}), we have:
\be  \Omega = a_2 + b_2 \ast. \lb{55} \ee
This ansatz leads only to the topological terms of gravity in the
total action, which we temporarily won't consider. Then, in
contrast to (\ref{33}), we have considered the ansatz of the
following type, which also is allowed:
$$  {{{\Omega_{\mu\nu}}^{\varphi\chi}}\,^{ab}}_{cd} =
a\,{\epsilon_{\mu\nu}}^{\alpha\varphi}(e_{\alpha}^a)
(e_{\beta}^b)(e^{\chi}_c) (e^{\beta}_d) +
b\,{\epsilon_{\mu\nu}}^{\varphi\chi}(e_{\alpha}^a)
(e_{\beta}^b)(e^{\alpha}_c)(e^{\beta}_d)$$ \be +
c\,{\epsilon_{\mu\nu}}^{\alpha\beta}(e_{\alpha}^a)
(e_{\beta}^b)(e^{\varphi}_c)(e^{\chi}_d),
 \lb{56} \ee
and
 \be {{{\Omega_{\mu\nu}}^{\varphi\chi}}\,^{a\tilde m}}_{c\tilde n} = (a + b +
c){\epsilon_{\mu\nu}}^{\varphi\chi}\delta^{a\tilde m}_{c\tilde n},
 \lb{57} \ee
\be {{{\Omega_{\mu\nu}}^{\varphi\chi}}\,^{\tilde m\tilde
n}}_{\tilde k \tilde l} = (a + b +
c){\epsilon_{\mu\nu}}^{\varphi\chi}\delta^{\tilde m \tilde
n}_{\tilde k \tilde l}.
 \lb{58} \ee
Now we are ready to consider the action which is a consequence of
our graviweak unification.

\subsection{The action of the graviweak unified model with \\ the renormalizable
gravity in the left and right worlds}

Using all our notations and equations (\ref{40})-(\ref{44}), we
can write:
 \be  F_{\mu\nu}^{ab} = \frac 12 (\omega_{\mu\nu}^{ab} - \frac 18
 \Sigma_{\mu\nu}^{ab}\phi^2),  \lb{59} \ee
where
 \be  \omega_{\mu\nu}^{ab} = \partial_{\mu}\omega_{\nu}^{ab} -
 \partial_{\nu}\omega_{\mu}^{ab} + \frac 12
 [\omega_{\mu}^{ac},\omega_{\nu}^{cb}],
\lb{60} \ee
and
 \be  \Sigma_{\mu\nu}^{ab} = e^a\wedge e^b
\lb{61} \ee
with the "metricity constrain"\,:
 \be \Sigma^a\wedge \Sigma^b =
\frac 13 \delta^{ab}\Sigma_c\wedge \Sigma^c. \lb{61a} \ee
Then
 \be  F_{\mu\nu}^{a\tilde m} = \frac 14 T_{\mu\nu}^a \phi^{\tilde
 m}
 - \frac 18 (e_{\mu}^a{\cal D}_{\nu}^{\tilde m \tilde n}
 -  e_{\nu}^a{\cal D}_{\mu}^{\tilde m \tilde n})\phi^{\tilde n},   \lb{62} \ee
where
 \be  T_{\mu\nu}^a = \partial_{\mu} e_{\nu}^a - \partial_{\nu}
 e_{\mu}^a + \frac 12 [\omega_\mu^{ab}, e_\nu^b] \lb{63} \ee
is a torsion, and
 \be  F_{\mu\nu}^{\tilde m \tilde n} = {F_W}_{\mu\nu}^{\tilde m \tilde n}
 \lb{64} \ee
is the curvature of the weak gauge field.
The action (\ref{38}) was calculated, using Eq.~(\ref{38a}) with
ansatz expressions (\ref{56})-(\ref{58}). Also
Eqs.~(\ref{59})-(\ref{64}) were used. The Eq.~(\ref{38a}) allows
us to return to the GR formalism, when the dynamics is described
by the metric tensor $g_{\mu\nu}$. Ignoring the fermionic matter,
we have no source for torsion, and the torsion (\ref{63}) is
absent in the action. Then the result is given by the following
integral:
 $$  I_{(GW)}= - \frac 1{4g_{uni}}\int_{\mathfrak M}
 d^4x \sqrt{-g}[\frac{(a+b+c)}{8}(\frac 12R|\phi|^2 - \frac 34|\phi|^4) + \frac
 1{16}(aR_{\mu\nu}R^{\mu\nu} + bR^2 +
 cR_{\mu\nu\varrho\sigma}R^{\mu\nu\varrho\sigma})$$ \be + (a+b+c)(\frac 12{\cal
 D}_{\mu}\phi{\cal
 D}^{\mu}\phi + \frac 14 {F_W}^i_{\mu\nu}{F_W}^{i\,\mu\nu})].  \lb{65} \ee
Considering the Friedmann-Lemaitre-Robertson-Walker (FLRW) homogeneous and isotropic metric,
we assume that the Gauss-Bonnet topological action
term vanishes, i.e. $c=0$, because the metric is conformally flat \ct{40sr}.
Here we want to emphasize, that the graviweak action of Ref.~\ct{1}
corresponds to the case $a=b=0$ and $c=3/2$.

However, in the present paper the action is considered in terms of the following expression:
 $$   I_{(GW)}= - \frac 1{4g_{uni}}\int_{\mathfrak M}
 d^4x \sqrt{-g}[\frac{w}{8}(\frac 12R|\phi|^2 - \frac 34|\phi|^4) $$ \be  + \frac
 1{16}(aR_{\mu\nu}R^{\mu\nu} + bR^2) + w(\frac 12{\cal
 D}_{\mu}\phi^\dag{\cal
 D}^{\mu}\phi + \frac 14 {F_W}^i_{\mu\nu}{F_W}^{i\,\mu\nu})],  \lb{66} \ee
where  $w=a+b$.

The parameters $w,a,b$  are coupling constants of the higher derivative gravity.

Assuming that at the first stage of the evolution (before inflation) the Universe
had the de Sitter spacetime, which is a maximally symmetric
Lorentzian manifold with a constant and positive background scalar curvature
$R_0$, we obtain a nontrivial vacuum solution to the action
(\ref{66}). This de Sitter spacetime has a non-vanishing Higgs vacuum expectation value (VEV):
\be  \phi_0^2 \equiv <\phi_0^2> = v^2 = \frac 13 R_0, \lb{67} \ee
��� $v = <\phi_0>$ is the vacuum expectation value (VEV)). The small fluctuations near
this vacuum expectation value are given by the fields
$\eta$:
\be    \phi^m = v + \eta^m.  \lb{68} \ee
As a result, we obtain the following action:
 $$     I_{(GW)} = - \frac 1{4g_{uni}}\int_{\mathfrak M}
 d^4x \sqrt{-g}[\frac{wv^2}{8}\big(\frac{R}{2}(1 + |\eta|^2/v^2)
 - \Lambda_0(1 + |\eta|^2/v^2)^2\big) $$ \be   + \frac {1}{16}(aR_{\mu\nu}R^{\mu\nu}
 + bR^2) + w(\frac 12{\cal
 D}_{\mu}\eta^\dag{\cal D}^{\mu}\eta + \frac 14 {F_W}^i_{\mu\nu}{F_W}^{i\,\mu\nu})],
\lb{69} \ee
in which the parameters $a,b,w$ are "bare"\, coupling constants corresponding to the Planck scale.

Here we can introduce the following relations:\\

1) the "bare" cosmological constant (the contribution of the gravitational zero modes) is:

 \be    \Lambda_0 = \frac 34 v^2 = \frac 14 R_0; \lb{70} \ee

2) the squared coupling constant of the weak interactions $g_W\equiv g_2$ at the Planck scale is:
 \be   g_W^2 = \frac {4g_{uni}}{w}; \lb{71} \ee

3) the Newton constant $G_N$ and the reduced Planck mass are:
\be   (M^{red.}_{Pl})^2 = {(8\pi G_N)}^{-1} = \frac{1}{\kappa^2} = \frac{wv^2}{32g_{uni}} = \frac {v^2}{8g_{W}^2}.\lb{72} \ee

Considering the running constant $\alpha_2^ {-1}(\mu)$, where $\alpha_2=g_2^2/4\pi$,
it is possible to make an extrapolation of this value to
the Planck scale \cite{6mp,7mp}, which gives the following result:
\be \alpha_2 (M_{Pl}) \sim 1/50, \qquad g_{uni}\sim 0.1. \lb{73} \ee
Having substituted in Eq.~(\ref{72}) the values of $g_W$ and $G_N=1/8\pi {(M_{Pl}^{red.})}^2$,
where $M_{Pl}^{ red.}\approx 2.43\cdot 10^{18}$
GeV, it is easy to obtain the VEV's value $v$, which in this case is located near
the Planck scale:
\be v=v_2\approx 3.5\cdot 10^{18} {\rm GeV}. \lb{73a} \ee

Then we have the following OW action near the Planck scale:
\be    I_{(OW)} = \int_{\mathfrak M} d^4x\sqrt{-g}[L_{(GW)} + L_{U(4)}], \lb{74} \ee
where the graviweak unification is described by the part of the total action  $I_{(OW)}$:
 $$    I_{(GW)} = - \int_{\mathfrak M} d^4x\sqrt{-g}L_{(GW)} \simeq - \int_{\mathfrak M}
 d^4x \sqrt{-g}[\frac{3(M^{red.}_{Pl})^2}{4\Lambda_0}\big(\frac{R}{2}|\phi|^2
 - \frac 34|\phi|^4\big ) $$ \be
 + \frac {1}{64g_{uni}}(aR_{\mu\nu}R^{\mu\nu} + bR^2) + \frac{1}{g_W^2}(\frac 12{\cal
 D}_{\mu}\phi^\dag{\cal D}^{\mu}\phi + \frac 14 {F_W}^i_{\mu\nu}{F_W}^{i\,\mu\nu})].
 \lb{75} \ee
Spontaneous symmetry breaking of the $Spin(4,4)$-invariant action of the graviweak unification
by non-trivial vacuum expectation values gives
the following actions in the ordinary and mirror worlds:
 $$  I^{(')}_{(GW)}= - \frac 1{4g_{uni}}\int_{\mathfrak M}
 d^4x \sqrt{-g^{(')}}[\frac{w^{(')}}{8}(\frac 12R^{(')}{|\phi^{(')}|}^2
 - \frac 34{|\phi^{(')}|}^4) + \frac
 1{16}(a^{(')}R^{(')}_{\mu\nu}{R^{(')}}^{\mu\nu}
 + b^{(')}{R^{(')}}^2  $$ \be
 + w^{(')}(\frac 12{\cal
 D}_{\mu}{\phi^{(')}}^{\dagger}{\cal D}^{\mu}\phi^{(')} +
 \frac 14 {F_{W^{(')}\mu\nu}^{(')i}}{F^{(')i}_{W^{(')}}}^{\mu\nu})],
   \lb{76} \ee
with $w^{(')}= a{(')} + b{(')}$. Here we assume the equality of the graviweak unification parameters: $g'_{uni}= g_{uni}$, which is a
consequence of the assumption that TOE existed at the early stage of the Universe, when MP was unbroken.

\section{Super-renormalizable gravity and the problem of unitarity}

The development of a quantum field theory of the Einstein-Hilbert GR faced a serious problem: quantum gravity based on the GR is
non-renormalizable, the traditional methods of renormalization cannot be used to eliminate the ultraviolet divergences appearing in its
perturbation theory.

Perturbative quantum gravity is a quantum theory of a spin two particle on a fixed background. Starting from the Einstein-Hilbert action
(\ref{23}), we introduce a splitting of the metric in a background part plus a fluctuation:
\be g_{\mu\nu} = g^0_{\mu\nu} + \kappa h_{\mu\nu},  \lb{77} \ee
where $h_{\mu\nu}$ is a graviton. Then we expand the action in powers of the fluctuation $\kappa h_{\mu\nu}$ around the fixed background
$g^0_{\mu\nu}$ (it may be Minkowski metric $\eta_{\mu\nu}$).

In the four-dimensional gravitational theory the superficial degree of divergence $D = 2+2L$ increases with the number $L$ of loops and thus we
are forced to introduce an infinite number of counter terms, i.e. an infinite number of coupling constants. This circumstance makes the theory
unpredictable.

An alternative way to quantize gravity is the introducing of some higher derivative terms into the classical action, treating them along with
other lower-derivative (Einstein-Hilbert and cosmological) terms. Such a theory of gravity was developed in \\
Refs.~\ct{40sr,1sr,2sr,3sr,4sr,5sr,6sr,8sr,9sr,11sr,12sr,14sr, 15sr,16sr,17sr,19sr,20sr,21sr,22sr}. For example, adding generic fourth order
derivative terms, one modifies propagator and vertices in such a way that the new quantum theory is renormalizable \ct{1sr,5sr}. This nice
property leads to establish the asymptotic freedom in the UV limit \ct{3sr,5sr}. Including new terms with derivatives higher than four we obtain
super-renormalizable theories of quantum gravity \ct{14sr,15sr,16sr,17sr,19sr,20sr,21sr}.

A first revolution in quantum gravity was introduced by Stelle \ct{1sr}, who suggested the following action of the higher derivative
gravitational theory:
\be S = - \int d^4x\sqrt{-g}[\alpha R_{\mu\nu} R^{\mu\nu} + \beta R^2 + \frac{1}{2\kappa^2}R]. \lb{78} \ee
This theory is renormalizable, but unfortunately contains a physical ghost (state of negative norm)  breaking unitarity in the theory.

A problem of ghosts is a very serious and paramount task in the theory of quantum gravity. Recently a new gravitational theory was suggested,
which is an approach to the quantum gravity. We can resume as follows the theoretical and observational consistency requirements for a full
theory of quantum gravity: (i) classical solutions must be singularity-free; (ii) Einstein-Hilbert action should be a good approximation of the
theory at a much smaller energy scale than the Planck mass; (iii) the spacetime dimension has to decrease with the energy in order to have a
complete quantum gravitational theory in the ultraviolet regime; (iv) the theory has to be perturbatively renormalizable at the quantum level;
(v) the theory has to be unitary, with no other pole in the propagator in addition to the graviton; (vi) spacetime is a single continuum of
space and time, and in particular, the Lorentz invariance is not broken, consistently with observations.

In Refs.~\ct{40sr,14sr,15sr,16sr,17sr,19sr,20sr} the Stelle theory was generalized to restore unitarity.  They considered a modification of the
Feynman rules where the coupling constants $g_i$ are no longer constant, but function of the momentum p. For particular choices of $g_i(p)$ and
$G_N(p)$ (gauge coupling constants and gravitational constant as functions of the 4-momenta), the propagators do not show any other pole above
the standard particle content of the theory, therefore the theory is unitary. On the other hand the theory is also finite if the coupling
constants go sufficiently fast to zero in the ultra-violet limit.

Super-renormalizable gravity (SRG) suggested in Refs.~\ct{40sr,14sr,15sr,16sr,17sr,19sr,20sr} is well defined perturbatively at the quantum
level. The corresponding gravitational Lagrangian is a �non-polynomial� extension of the renormalizable quadratic Stelle theory \ct{1sr} and it
has the following general structure:
\be L = \frac{\alpha_k}{2\kappa^2} R + \alpha R_{\mu\nu}R^{\mu\nu} + \beta R^2
 + R_{\mu\nu}h_2(-\Box_{\Lambda})R^{\mu\nu}
+ Rh_0(-\Box_{\Lambda})R, \lb{79} \ee
where $\Box$ is the covariant $\rm {D^{'}}$Alembertian operator, $\Box_{\Lambda}:=\Box/{\Lambda}$, and $\Lambda$ is an invariant mass scale; the
two functions $h_i(-\Box_{\Lambda})$ (i=0,2) are not polynomial but entire, i.e. without poles or essential singularities; $\alpha_k, \alpha,
\beta$ are coupling constants subjected to quantum renormalizations.

Thus, there was introduced a nonlocal extension of the higher-derivative gravity, which is perturbatively renormalizable and unitary in any
dimension D. The four-dimensional theory is easily obtained from the Stelle theory \ct{1sr} by introducing in the action two entire functions,
i.e. form factors, between the Ricci scalar square and the Ricci tensor square :
$$R^2 \to R h_0(\Box_{\Lambda})R,$$ \be
R_{\mu\nu}R^{\mu\nu}\to R_{\mu\nu}h_2(\Box_{\Lambda})R^{\mu\nu}. \lb{80} \ee
The main reason for introducing the entire functions $h_i(z)$ (i=0,2) is to avoid ghosts  and any other new pole in the graviton propagator.

The gravitational  theory constructed in Refs.~\ct{40sr,14sr,15sr,16sr,17sr,19sr,20sr,21sr,22sr} is renormalizable at one loop and finite from
two loops on. Since only a finite number of graphs are divergent, then the theory is super-renormalizable.

Our graviweak unification theory suggests the action (\ref{75}), in which a gravitational part of theory is super-renormalizable and
asymptotically-free, but is not unitary in flat-space perturbation theory. With aim to make this theory unitary, it is necessary to introduce a
nonlocal extension of the higher-derivative gravity, given by the procedure (\ref{80}). Nevertheless, we can consider the transplanckian running
of the "coupling constants" $w,\,a,\,b$ given by the action (\ref{75}) with aim to show the asymptotic freedom of this theory.

\subsection{Running constants in the super-renormalizable gravity
 predicted by the graviweak unification}

The study of the renormalization group flow of higher derivative gravity is based on the Schwinger-DeWitt technique, generalized by Barvinsky
and Vilkovisky \ct{26sr}. The one-loop renormalization constants in higher-derivative gravity were first calculated by Julve and Tonin \ct{3sr}.
The final correct result was obtained in Refs.\ct{15sr,21sr} (see also \ct{22sr}). Using this result, we can calculate the gravitational
functional renormalization group equations (FRGEs) for parameters $w,\,a,\,b$ in the action (\ref{75}) of the graviweak unification.

It is convenient to consider the running of coupling constants:
\be \gamma=\frac{w}{64g_{uni}},\quad \tilde a = \frac{a}{64g_{uni}},\quad \tilde b = \frac{b}{64g_{uni}}, \lb{80a} \ee
given by the RGEs:
\be \frac{d\gamma}{dt}=\beta_1,\quad \frac{d\tilde a}{dt}=\beta_2,\quad \frac{d\tilde b}{dt}=\beta_3, \lb{80b} \ee
when $t=\ln(\mu/\mu_0)$, and $\mu$ is a scale of energy. We choose the renormalization point $\mu_0=M_{Pl}^{red.}$.

In the one-loop approximation we obtain the following results \ct{15sr,21sr,22sr}:
\be \gamma(t)=\gamma(\mu_0) + \beta_1t, \quad
 {\tilde a}(t)={\tilde a}(\mu_0) + \beta_2t,\quad
 {\tilde b}(t)={\tilde b}(\mu_0) + \beta_3t,  \lb{81a} \ee
where
\be \beta_3\approx \beta_2=\frac {133}{10}, \quad {\tilde b}(M_{Pl}^{red.})=-{\tilde a}(M_{Pl}^{red.})=
\frac1{12}\left(\frac{M_{Pl}^{red.}}{m}\right)^2,
                               \lb{81b} \ee
according to the action (\ref{75}).

For the dimensionless coupling constant $16\pi G_N\Lambda=\gamma^{-1}$ we obtain (see Ref.~\ct{22sr}):
\be \gamma(t)= \gamma(M_{Pl}^{red.}) + \frac{1+4\omega_1}{6}\beta_2 t,
 \lb{81c} \ee
where  $\gamma(M_{Pl^{red.}})\approx 0.32$ is given by the relations (\ref{80a}) and (\ref{70})-(\ref{73a}) ; $\omega_1\approx -0.02$ is a fixed
point \ct{22sr} (in general, theory has two fixed points, see \ct{22sr}). Finally, we obtain:
\be \lim _{t\to \infty}\gamma(t)\simeq 0.32 + 2.04t,
 \lb{82} \ee
what means that the dimensionless gravitational coupling constant $G_N\Lambda$ ($=1/16\pi \gamma$) is asymptotically free (here $\Lambda\equiv
\Lambda_0$, which is a bare cosmological constant determined only by zero modes of gravitational fields).

The next development of this theory is given in the Part II of our investigation, where we 
consider algebraic spinors of the standard four-dimensional Clifford algebra with a left-right symmetry and imagine the embedding of the
fermion families into the groups $U(4)_{L,R}$ with a final formation of the SM,SM'-groups of symmetry in the OW and MW, respectively. Then we
consider the inflation model, predicted by our theory of the  graviweak unification, and  the Multiple Point Model (MPM), assuming the
existence of several minima of the Higgs effective potential with the same energy density. We show that MPM is in agreement with our graviweak
unification. The predictions of the top-quark and Higgs masses are given from the assumption that there exist two vacua into the SM: the first
one -- at the Electroweak scale ($v_1\simeq 246$ GeV), and the second one -- at the Planck scale ($v=v_2\sim 10^{18}$ GeV).

\section{Summary and Conclusions}

~~~~~1. In the present paper we constructed a theory of the
unification of super-renornalizable gravity with weak $SU(2)$ gauge
and Higgs fields.

2. We have given arguments that recent astrophysical and cosmological measurements lead to a model of the Mirror World with broken Mirror Parity
(MP), in which the Higgs VEVs of the visible and invisible worlds are not equal: $\langle\phi\rangle=v, \quad \langle\phi'\rangle=v' \quad {\rm
{and}}\quad v\neq v'$. The parameter characterizing the violation of the MP is $\zeta = v'/v \gg 1$. We have used the estimate $\zeta \sim 100$,
in accordance with Refs.~\ct{5y,2y,3y,4y}. In this model, we showed that the action for gravitational and $SU(2)$ Yang--Mills and Higgs fields,
constructed in the ordinary world (OW), has a modified duplication for the hidden (mirror) sector of the Universe (MW).

3. We discussed the problems of communications between visible and
invisible worlds. Mirror particles have not been seen so far in
the visible world, and the communication between visible and
hidden worlds is hard. This communication is given by the
$L_{(mix)}$-term of the total Lagrangian of the Universe.

4. We started with an extended $\mathfrak g =
\mathfrak{spin}(4,4)_L$-invariant Plebanski action in the visible
Universe, and with $\mathfrak g =
\mathfrak{spin}(4,4)_R$-invariant Plebanski action in the MW.

5. We showed that the graviweak symmetry breaking leads to the
following subalgebras: $\tilde {\mathfrak g} = {\mathfrak
sl}(2,C)^{(grav)}_L \oplus {\mathfrak su}(2)_L$ -- in the ordinary
world, and $\tilde {\mathfrak g}' = {{\mathfrak
sl}(2,C)'}^{(grav)}_R \oplus {\mathfrak su}(2)'_R$ -- in the
hidden world. These subalgebras contain the self-dual left-handed
gravity in the OW, and the anti-self-dual right-handed gravity in
the MW.

6. We developed a graviweak unification model in both, left-handed
and right-handed (visible and invisible) sectors of the Universe.
We considered the left and right worlds OW and MW, existing at the
first stage of the Universe described by the symmetries
$SL(2,C)^{(grav)}_L\times SU(2)_L\times U(4)_L$ and
$SL(2,C)^{(grav)}_ R\times SU(2)_R\times U(4)_R$, respectively.

7. In contrast to Refs.~\ct{1} and \ct{2}, we considered a general class of solutions for Ansatz, which gives any non-trivial values of
parameters introducing the graviweak unification action. As a result, we propose a class of the super-renormalizable (finite) theory of gravity,
providing an ultraviolet completion of the gravitational theory. This class of theory has a generalization, searching for the unitary,
asymptotically-free and perturbatively consistent theory of quantum gravity. We have shown that the self-consistent graviweak unification is
described by the higher-derivative super-renormalizable  gravity, and this graviweak unification exists only at the high (Planck) scale.

8. The nontrivial vacuum solutions corresponding to the obtained actions are non-vanishing Higgs vacuum expectation values (VEVs):
$v^{(')}=\langle\phi^{(')}\rangle=\phi^{(')}_0$, which are not equal for the visible and mirror (or hidden) worlds.

9. Considering the graviweak unification,  we obtained after a symmetry breaking {Newton'}s constant $8\pi G_N = {64g_{uni}}/{3v^2}$, and the
bare cosmological constant $\Lambda_0 = \frac 34 v^2 = R_0/4$, where $v$ is given by the second vacuum of the effective Higgs potential, and
$R_0$ is a constant de Sitter spacetime background curvature.

10. We have calculated the graviweak action near the second local minimum of the effective Higgs potential, corresponding to the second vacuum
with the VEV $v=v_2\sim 10^{18}$ GeV.

11. We considered the renormalization group flow of the higher derivative gravity. We presented (in the 1-loop approximation) the running of the
super-renormalizable gravitational coupling constants, predicted by our graviweak unification model. We showed that the dimensionless
gravitational coupling constant $G_N\Lambda_0$) is asymptotically free (here $\Lambda_0$ is a bare cosmological constant determined only by zero
modes of gravitational fields).

\section{Acknowledgments}

We thank M. Chaichian for useful discussions. H.B.N. wishes to thank the Niels Bohr
Institute for the support. L.V.L. greatly
thanks the Niels Bohr Institute for hospitality  and
financial support. The support  of the Academy of Finland under the Projects No. 136539
and 272919 and of the Magnus Ehrnrooth Foundation is gratefully acknowledged.

\end{document}